\documentclass[12pt]{article}
\usepackage[export]{adjustbox}
\usepackage{amssymb, amsmath}
\usepackage{booktabs}
\usepackage[table,xcdraw]{xcolor}
\usepackage{multirow}
\usepackage{rotating,tabularx}
\usepackage{comment}
\usepackage[T1]{fontenc}
\usepackage{authblk}
\usepackage{color}

\usepackage{scicite}

\usepackage{times}

\newenvironment{sciabstract}{%
\begin{quote} \bf}
{\end{quote}}

\newcounter{lastnote}

\title{Low phase noise THz generation from a fiber-referenced Kerr microresonator soliton comb}

\author[1,2,*]{Naoya Kuse}
\author[3]{Kenji Nishimoto}
\author[1]{Yu Tokizane}
\author[3]{Shota Okada}
\author[4]{Gabriele Navickaite}
\author[4]{Michael Geiselmann}
\author[1,5]{Kaoru Minoshima}
\author[1]{Takeshi Yasui}

\affil[1]{Institute of Post-LED Photonics, Tokushima University, 2-1, Minami-Josanjima, Tokushima, Tokushima 770-8506, Japan}
\affil[2]{PRESTO, Japan Science and Technology Agency, 4-1-8 Honcho, Kawaguchi, Saitama, 332-0012, Japan}
\affil[3]{Graduate School of Science, Technology and Innovation, Tokushima University, 2-1, Minami-Josanjima, Tokushima, Tokushima 770-8506, Japan}
\affil[4]{LIGENTEC SA, EPFL Innovation Park L, Chemin de la Dent-d'Oche 1B, Switzerland  CH-1024 Ecublens, Switzerland}
\affil[5]{Graduate School of Informatics and Engineering, The University of Electro-Communications, 1-5-1 Chofugaoka, Chofu, Tokyo 182-8585, Japan}

\affil[*]{Corresponding author: kuse.naoya@tokushima-u.ac.jp}

\date{}

\begin{document} 


\baselineskip24pt


\maketitle

\begin{sciabstract}
Abstract\\
THz oscillators generated via frequency-multiplication of microwaves are facing difficulty in achieving low phase noise. Photonics-based techniques, in which optical two tones are translated to a THz wave through opto-electronic conversion, are promising if the relative phase noise between the two tones is well suppressed. Here, a THz ($\approx$ 560 GHz) wave with a low phase noise is provided by a frequency-stabilized, dissipative Kerr microresonator soliton comb. The repetition frequency of the comb is stabilized to a long fiber in a two-wavelength delayed self-heterodyne interferometer, significantly reducing the phase noise of the THz wave. A measurement technique to characterize the phase noise of the THz wave beyond the limit of a frequency-multiplied microwave is also demonstrated, showing the superior phase noise of the THz wave to any other photonic THz oscillators (\textgreater \ 300 GHz).
\end{sciabstract}


\section*{Introduction}
Spectral purity of electromagnetic fields, which is quantified by phase noise, is one of the most important parameters when the electromagnetic fields are used as oscillators. A distinguished example in the optical domain is ultra-stable, low-noise continuous-wave (CW) lasers for optical clockworks \cite{matei20171, beloy2021frequency}. In the microwave domain, low phase noise oscillators have been widely used both in the scientific and industrial world such as radars, communications, radio astronomy, and particle accelerators \cite{ghelfi2014fully, clivati2017vlbi, schulz2015femtosecond}. Recently, the demand for oscillators in the mm and THz ranges has been expanded mainly for next-generation communication (5G and 6G), including radars, molecular clocks, and wireless communications \cite{dang2020should, grajal2017compact, wang2018chip, koenig2013wireless}.

Optical frequency combs \cite{diddams2020optical} based on mode-locked fiber or solid-state lasers coherently connect the optical and microwave frequency through photodetection, where ultra-low phase noise of the repetition frequency ($f_{\rm rep}$: repetition frequency) is faithfully transferred to microwaves \cite{fortier2011generation, xie2017photonic, nakamura2020coherent}, reaching the shot noise floor of -170 dBc/Hz at the 1-kHz frequency offset with $1/f$ ($f$: frequency offset from a carrier) slope for a 12 GHz carrier \cite{xie2017photonic}. However, fiber combs and solid-state laser combs are not suitable for the generation of mm and THz waves owing to the inherently narrow comb mode spacings (\textless \ 1 GHz). On the other hand, other photonics-based methods such as optoelectronic oscillators (OEOs) \cite{yao1996optoelectronic, eliyahu2008phase} and heterodyning of two CW lasers \cite{quraishi2005generation, ducournau2011highly, kuse2018photonic, li2019low, kittlaus2021low} are suitable for the generation of mm and THz waves. In particular, heterodyne methods can be easily applied to a higher frequency. Among the heterodyne methods, two Brillouin lasers from a single cavity produce the small relative phase noise of the two CW lasers \cite{kuse2018photonic, li2019low}. Also, two self-injection locked lasers from high-Q cavities have been employed for low-phase-noise W-band (92 GHz) radar \cite{kittlaus2021low}.  

Owing to the advent of integrated optical frequency combs based on dissipative Kerr microresonator soliton combs (Kerr combs) \cite{Herr_soliton, kippenberg2018dissipative}, the generation of mm and THz waves from Kerr combs has been explored \cite{zhang2019terahertz, wang2021towards, tetsumoto2021optically} to outperform the above heterodyne methods using two CW lasers. The free spectral range (FSR) of microresonators used to generate Kerr combs can be easily more than 100 GHz, which is suitable for the generation of mm and THz waves through opto-electronic conversion by photodetectors. Moreover, Kerr combs are a mode-locked state, where a double balance of gain and loss as well as nonlinearity and dispersion exists, providing low-noise, coherent, ultra-short pulses. A mm wave (100 GHz) has been generated from a 100-GHz, free-running Kerr comb, showing a strong coherence between the comb modes \cite{wang2021towards}. To improve the phase noise, a 300-GHz Kerr comb is stabilized to two Brillouin lasers, where optical frequency division from 3.6 THz (= frequency separation of the two Brillouin lasers) to 300 GHz (= $f_{\rm rep}$ of the Kerr comb) is utilized, suppressing the phase noise by 40 dB in the wide range of frequency offsets for a 300-GHz carrier \cite{tetsumoto2021optically}. However, the method based on Brillouin lasers requires temperature stabilization (\textless \ 10-mK temperature stability), operation in vacuum, and additional two phase-locked loops to suppress the mode-hopping of the Brillouin lasers \cite{danion2016mode}. Also, two narrow-linewidth CW lasers are required in addition to the Kerr comb. Moreover, owing to the coupling between the laser dynamics and cavity fluctuation, the phase noise of the Brillouin laser at the low frequency offsets is unsatisfactory, limiting the phase noise of the 300-GHz wave at the low frequency offsets. Another method to stabilize the $f_{\rm rep}$ of Kerr combs is based on a two-wavelength delayed self-heterodyne interferometer (TWDI) \cite{kuse2019control, kwon2022ultrastable}, in which a long fiber in an imbalanced Mach-Zehnder interferometer (i-MZI) works as a reference. TWDI is the extension of i-MZIs with a long fiber, which has been used to stabilize a CW laser \cite{kefelian2009ultralow} and comb modes of a $f_{\rm ceo}$-stabilized frequency comb \cite{kwon2020generation}, to the two different wavelengths, enabling the detection of the phase noise of $f_{\rm rep}$ \cite{jung2015all}. In the method, the phase noise of the $f_{\rm rep}$ is measured in a self-referencing way owing to the time delay between the two arms in the i-MZI, allowing simple and robust operation of the stabilization system. Furthermore, the phase noise limit at the frequency offsets below 10 kHz set by the fiber length fluctuation can be minimized \cite{kwon2017all, kwon2020generation}, compared to the system using a Kerr comb and two Brillouin lasers \cite{tetsumoto2021optically}. However, despite the importance of the THz oscillator and advantages of the TWDI over the system using a Kerr comb and two Brillouin lasers \cite{tetsumoto2021optically}, TWDIs have not been utilized for THz waves generation from a Kerr comb, yet.


In this report, we demonstrate the generation of a low phase noise THz ($\approx$ 560 GHz) wave from a $f_{\rm rep}$ ($\approx $ 560 GHz) stabilized Kerr comb. The low phase noise is obtained by stabilizing the THz Kerr comb to a TWDI, where the phase noise of the $f_{\rm rep}$ is extracted by mixing the phase noise of the two comb modes at different frequencies and feeding-back to the pump CW laser. The stabilization suppresses the phase noise of the $f_{\rm rep}$  by more than 40 dB in the wide range of frequency offsets. Due to the lack of established methods to characterize carriers with THz frequency and low phase noise, we also propose and develop a method for the measurement of the phase noise of the THz wave beyond the limit of conventional microwave technologies, where two low noise THz waves to be characterized are down-converted to a microwave at a Schottky barrier diode (SBD) by the square-law detection. With the generation and measurement of the low noise THz wave, we show that the phase noise of the stabilized $f_{\rm rep}$ in the THz regime is transferred to the THz wave, exhibiting the superior phase noise of the THz wave to any other photonic THz (\textgreater \ 300 GHz) oscillators.

\section*{Results}
\subsection*{Architecture for the THz generation}
A basic architecture for the generation of a low phase noise THz is shown in Fig. 1. There are three main elements: (1) Kerr comb, (2) detection and stabilization of the $f_{\rm rep}$, and (3) THz wave generation. A Kerr comb is generated from a high-Q Si$_3$N$_4$ microresonator. The phase noise of the $f_{\rm rep}$ is detected in the optical domain by using a TWDI and referencing a long fiber in the TWDI as explained in the following (see a photo in the supplementary material). The detected phase noise of the $f_{\rm rep}$ works as an error signal for a feedback loop, which is closed by feeding back to the Kerr comb. After suppressing the phase noise of the $f_{\rm rep}$, two comb modes of the Kerr comb are extracted by an optical filter, followed by a uni-traveling-carrier photodiode (UTC-PD). The optical filter is required to avoid the destructive interference between the comb modes unless the dispersion of the Kerr comb is compensated. The UTC-PD converts the optical two tones to a THz wave, whose frequency corresponds to the spacing of the two tones. The phase noise and frequency stability of the THz wave inherit those of the $f_{\rm rep}$ of the Kerr comb, i.e. relative phase noise and frequency stability between the two comb modes, providing the generation of a low phase noise THz wave from the $f_{\rm rep}$-stabilized Kerr comb.

\subsection*{Experimental setup}
Figure 2(a) shows an experimental setup not only to generate a low phase noise THz wave but also to measure the phase noise and frequency stability of the THz wave. Two Kerr combs (combs 1 and 2) are generated by using two independent systems (detail is shown in the method section, and the supplementary material), in which a single-longitudinal CW laser (external cavity diode laser: ECDL) is coupled into a high-Q Si$_3$N$_4$ microresonator \cite{pfeiffer2018photonic}. The Kerr comb passes through a bandstop filter to reject the residual pump CW laser. The repetition frequencies of combs 1 and 2 ($f_{\rm rep1}$ and $f_{\rm rep2}$) are approximately 560 GHz. The Kerr comb is split into two. One is directed to a TWDI \cite{kuse2019control}, and the other is used for THz generation. TWDI consists of an i-MZI with two outputs followed by optical bandpass filters. A large time delay between the two arms in the i-MZI is employed for the self-referencing by installing a long fiber ($\approx$ 20, 80, 100, 110 m in the experiments). An acousto-optic modulator (AOM) is also installed in the i-MZI for the self-heterodyne detection. The optical bandpass filters (OBPFs) select a pair of comb modes at a specific frequency ($n$ and ($n+\Delta$) th in Fig. 2(b)), which are directed into photodetectors, generating delayed-self heterodyne signals. More details of the experimental setup of the TWDI are shown in the method section and supplementary material. The delayed self-heterodyne signals show the phase noise PSDs of the $n$th comb mode ($|H(if)|^2\{L_{\rm ceo}(f) + n^2L_{\rm rep}(f)\}$) and the ($n + \Delta$) th comb mode ($|H(if)|^2\{L_{\rm ceo}(f) + (n+\Delta)^2L_{\rm rep}(f)\}$). Here, $H(if)$, $L_{\rm ceo}(f)$, and $L_{\rm rep}(f)$ are the transfer function of the i-MZI (see in the method section) and phase noise PSDs of the carrier-envelope offset frequency and $f_{\rm rep}$ of the Kerr comb, respectively. By mixing the two delayed self-heterodyne signals, a DC signal with the phase noise PSD of $|H(if)|^2\cdot \Delta^2 L_{\rm rep}(f)$ is obtained, while canceling out $L_{\rm ceo}(f)$ owing to the common-mode rejection. Since the DC signal contains $L_{\rm rep} (f)$ only with coefficients of $H(if)$ and $\Delta$, a feedback loop to make the DC signal fixed at zero reduces $L_{\rm rep}(f)$. Throughout the experiments in this report, $\Delta$ is five (the +1st and -4th for comb 1 and +2nd and -3rd for comb 2). In the experiments, the loop is closed by feeding back to the pump current of the pump CW laser and PZT in the pump CW laser cavity for fast and large-range control, respectively, modulating the frequency of the pump CW laser.
To characterize the THz wave, two comb modes from combs 1 and 2 (four comb modes in total) are extracted by an optical filter (4000s from Finisar) (Fig. 2(c), top left) and input into a UTC-PD with an integrated antenna (IOD-PWAN-13001-2 from NTT Electronics). THz waves with the carrier frequencies of $f_{\rm rep1}$ and $f_{\rm rep2}$ (Fig. 2(c), top right) propagate in free space. The two THz waves are focused onto a SBD (WR1.9ZBD from Virginia Diodes Inc.), generating a microwave with the carrier frequency of $|f_{\rm rep1} - f_{\rm rep2}|$ ($\approx$ 1.26 GHz) (Fig. 2(c), bottom) through the square-law detection. 
The phase noise and frequency stability of the microwave are the sums of those of the two THz waves, which inherit the phase noise and frequency stability of the $f_{\rm rep}$ of the two Kerr combs. However, owing to the use of a single UTC-PD for the generation of the two THz waves, the correlated phase noise between the two THz waves originated in the opto-electronic conversion process at the UTC-PD such as the amplitude-to-phase noise conversion and flicker noise cannot be included. Therefore, the phase noise characterization presented here shows the lower limit of the phase noise of the THz wave. Alternatively, the two Kerr combs can be directed to two independent UTC-PDs and downconverted by a fundamental mixer to cover the phase noise arising in the photodetection process. Owing to the availability of the UTC-PD in our lab, we use a single UTC-PD.
More details about the generation and characterization of the THz wave are shown in the method section and supplementary material. 

\subsection*{Experimental results}
Figures 3(a) and (c) shows the optical spectra of combs 1 and 2 before the bandstop filters. The smooth sech$^2$ envelope with the comb mode spacing equal to the FSR of the microresonator indicates the Kerr comb is a single soliton state. After rejecting the strong residual pump CW laser by the bandstop filter, the Kerr combs are amplified (not shown in Fig. 2(a)) by an erbium-doped fiber amplifier (EDFA) for the following experiments as shown in Figs 3(b) and (d) for combs 1 and 2, respectively. The optical signal-to-noise ratio (OSNR) of the comb modes is significantly degraded by the amplified spontaneous emission (ASE) during the amplification down to about 28 dB at the resolution bandwidth (RBW) of 1 nm for the comb modes. The OSNR is one of the limiting factors of the sensitivity of the TWDI, which is further discussed later and in the supplementary material. Figure 3(e) shows the comb modes used for the THz waves generation. The optical power of each comb mode is about 8 mW, and 32 mW in total is input into the UTC-PD. The THz waves are down-converted to a microwave by mixing the two THz waves at the SBD by the square-law detection. The RF spectrum of the microwave is shown in Fig. 3(f). The SNR of the microwave is about 58 dB at the RBW of 100 kHz, resulting in a measurement noise floor of -108 dBc/Hz.

The repetition frequencies of combs 1 and 2 are stabilized to a long fiber in TWDIs 1 and 2. The imbalanced fiber length for TWDIs 1 and 2 are 100 m and 80 m, respectively. The fiber length is selected such that the phase noise around the 10-kHz frequency offset is optimized.
RF spectra of the down-converted microwave are shown in Fig. 4(a). When the feedback loops both for combs 1 and 2 are closed, the phase noise close to the carrier is suppressed, showing the significant narrowing of the linewidth. A servo bump is observed around the 200-kHz frequency offset. 
The single-sideband (SSB) phase noise PSD of the down-converted signal ($L_{\rm mw}(f)$) is shown in Fig. 4(b) (red curve), which is measured by an electric spectrum analyzer (N9030A from Agilent Technologies). Compared with the phase noise PSDs of the $f_{\rm rep}$ of free-running combs 1 (blue in Fig. 4(b)) and 2 (green in Fig. 4(b)), the phase noise is suppressed by more than 40 dB at the frequency offsets below 10 kHz, reaching -49, -77, -99, -95, and -108 dBc/Hz at the frequency offsets of 100 Hz, 1 kHz, 10 kHz, 100 kHz, and 1 MHz, respectively. The integrated timing jitter from 100 Hz to 1 MHz can be estimated as 14 fs from the phase noise PSD. Note that $L_{\rm mw}(f)$ is the sum of the phase noise PSDs of $f_{\rm rep1}$ and $f_{\rm rep2}$ ($L_{\rm rep1}(f)$ and $L_{\rm rep2}(f)$), indicating that the phase noise of a single THz wave is better than the shown phase noise. 
The obtained phase noise is compared with literature reporting the generation of \textgreater \ 100 GHz waves from Kerr combs in Fig. 6. For a fair comparison, the phase noise scaled to 560 GHz is also shown. The phase noise of the $f_{\rm rep}$ of the free-running 560-GHz Kerr comb (comb 1) is much better than that of the 100-GHz Kerr comb when scaling to 560 GHz is considered \cite{wang2021towards}. Smaller free-running phase noise is important to achieve the lower phase noise with the stabilization because of the limited feedback gain. When stabilized, our result is compared with the system using a Kerr comb and two Brillouin lasers \cite{tetsumoto2021optically}. Our phase noise is slightly better at the frequency offsets of \textgreater \ 10 kHz. At the frequency offsets below 10 kHz, which is essential for e.g. Doppler radar \cite{kittlaus2021low}, our results show a superior (more than 20 dB better) performance with a simple aluminum enclosure for the TWDI. On the contrary, in the system using a Kerr comb and two Brillouin lasers \cite{tetsumoto2021optically}, despite a careful enclosure (100 mTorr in a vacuum chamber) for the Brillouin lasers, the obtained phase noise at the frequency offsets is limited by the references, i.e. relative phase noise between two Brillouin lasers, which might be difficult to further improve. Also, again note that our system is simpler and more robust as explained in the introduction. 

Not only the short-term stability (i.e. phase noise) but also the long-term stability is improved by locking the $f_{\rm rep}$ to the long fiber. Figure 4(c) shows the frequency drift of the down-converted microwave. Without the locking, the frequency drift of 6.6 MHz for one hour is observed, whilst the locking suppresses the frequency drift down to 120 kHz. In both cases, the linear drift would be caused by the temperature fluctuation (\textless \ 0.1 K in our lab for 1 hour). However, the influence is more significant ($\times$ 55) for the free-running Kerr comb, because the temperature fluctuation couples various factors such as fluctuations of FSR, detuning, and pump power. On the other hand, the frequency drift with the stabilization to the TWDI just follows the fiber length fluctuation if the feedback gain is large enough. When the linear drift is removed by signal processing, the standard deviation also shows an improvement from 390 kHz to 20 kHz. Fractional frequency instability is also evaluated from the results of the frequency drift (Fig. 4(d)). The fractional frequency instability is improved by two orders of magnitude, reaching 10$^{-10}$ at the 0.1-s averaging time, which is gradually degraded owing to the fluctuation of the fiber length in the TWDI. The frequency instability of $1.4 \times 10^{-11}$ at 1 ms averaging time is also estimated from the phase noise PSD in Fig. 4(b).

In Fig. 5(a), $L_{\rm mw}(f)$ is compared with the sum of the out-of-loop phase noise PSDs of the stabilized $f_{\rm rep1}$ and $f_{\rm rep2}$, which are measured by using the two TWDIs (see the supplementary material). Here, the in-loop phase noise PSD of the stabilized $f_{\rm rep}$ is defined as the phase noise PSD of the error signal divided by $\Delta ^2 |H(if)|^2$ of the TWDI, which is used for the stabilization. For the measurement of the out-of-loop phase noise PSD of $f_{\rm rep}$, another TWDI is used, and out-of-loop phase noise PSD of the stabilized $f_{\rm rep}$ is defined as the phase noise PSD of the error signal divided by $\Delta ^2 |H(if)|^2$ of the other TWDI. 
Although $L_{\rm mw}(f)$ overlaps with the sum of the out-of-loop phase noise PSDs of $f_{\rm rep1}$ and $f_{\rm rep2}$ at the wide frequency offsets, a deviation is observed at the frequency offsets below 300 Hz. The excess noise can be originated from the opto-electronic conversion at the UTC-PD. In addition, since a single UTC-PD is used, the excess phase noise may be underestimated owing to the correlation of the opto-electronic conversion process within the UTC-PD. More investigation of the excess noise such as the amplitude-to-phase noise conversion and flicker noise from the UTC-PD with THz bandwidth is left for future research.
Above the frequency offset of 400 kHz, $L_{\rm mw}(f)$ is limited by the measurement noise floor. 
In the following, we investigate the limit of the phase noise of the stabilized $f_{\rm rep}$ to indirectly investigate the limit of the phase noise of the THz wave. Figure 5(b) shows the out-of-loop (the blue curve) and in-loop (the red curve) phase noise of the stabilized $f_{\rm rep1}$. The out-of-loop phase noise of the stabilized $f_{\rm rep1}$ follows the in-loop phase noise at the frequency offsets above the 10 kHz. To decrease the phase noise at this range, Kerr combs with intrinsically small phase noise are required, which would be realized by operating in the quiet points \cite{yi2017single}, optimizing the dispersion of microresonators \cite{stone2020harnessing}, and using laser cooling techniques \cite{drake2020thermal, nishimoto2022thermal, lei2022thermal}. Alternatively, a shorter fiber can be used in the TWDI to increase the feedback gain, at the expense of the increase of the fiber noise limit and OSNR limit of the TWDI. Figure 5(c) shows the limit of the fiber noise. Although the thermal noise of the fiber is far below \cite{wanser1992fundamental, duan2012general, dong2016observation} (the gray curve in Fig. 5(c)), the fiber is influenced by acoustic noises. The fiber noise is prominent at the frequency offsets below 1 kHz. Currently, the i-MZI in the TWDI is just enclosed in an aluminum box with a thickness of 5 mm. To reduce the fiber noise, more delicate handling of the fiber such as putting into an enclosure with acoustic damping foam \cite{jung2015all} and low vacuum environment \cite{jeong2020ultralow} on vibration immune table, at the cost of complexity, is useful. Another method to reduce the fiber noise limit ($\propto L^{-1}$, $L$: fiber length difference in the i-MZI) is to employ a longer fiber. However, the use of a longer fiber decreases the feedback gain/bandwidth, increasing the in-loop limit. Figure 5(d) shows the OSNR limit of the TWDI, which is determined by the noise generated from the photodetection and subsequent electric components. At the photodetection, white noise is generated from the limited OSNR of the comb mode, which shows the OSNR limit with a $\frac{1}{f^2}$ slope after divided by $\Delta^2 |H(if)|^2$. The OSNR limit is influential at the frequency offsets below 10 kHz. To lower the OSNR limit, the OSNR of the comb mode that is used in the TWDI needs to be increased. Currently, the OSNR is limited by the ASE from the EDFA used after generating the Kerr comb. Overcoupled-microresonator would provide larger comb mode power \cite{riemensberger2020massively, jang2021conversion}, which reduces the ASE. Alternatively, the comb mode can be amplified by using injection locking, which provides more than 50-dB OSNR at the RBW of 1 nm \cite{kuse2022amplification}. In another way (or combined with injection locking), the use of largely separated comb modes also reduces the sensitivity limit owing to the larger factor of $\Delta$. With a fixed OSNR and $\Delta$, a longer fiber reduces the OSNR limit, however, at the cost of the reduction of the feedback gain/bandwidth, increasing the in-loop limit. Since both the fiber noise limit and OSNR limit would be improved by using a longer fiber, the better phase noise at the low frequency offset and the better frequency instability would be obtained until the in-loop limit is reached. More details for this paragraph are shown in the supplementary material.


\section*{Conclusions}
We have generated a low noise THz ($\approx$ 560 GHz) wave, which has been characterized by a measurement system for the low noise THz wave. The THz wave has been generated from a Kerr comb, whose $f_{\rm rep}$ has been stabilized to a long fiber in a TWDI, which allows the detection of the phase noise of the $f_{\rm rep}$ in a self-referenced way. To characterize the phase noise and frequency stability of the THz wave, we employed a measurement system, which eliminates the use of a frequency-multiplied microwave as a reference, thus, enabling the phase noise characterization beyond the limit of the conventional microwave technologies. 
With the generation and measurement of the low noise THz wave, we showed the SSB phase noise PSD of -49, -77, -99, -95, and -108 dBc/Hz at the frequency offsets of 100 Hz, 1 kHz, 10 kHz, 100 kHz, and 1 MHz, respectively. However, note that the measured phase noise is blind to the correlated excess noise between the two THZ waves arising in the opto-electronic conversion process at the UTC-PD, and further study of the excess noise of the UTC-PD with the THz bandwidth such as the amplitude-to-phase conversion and flicker noise is required.
According to an out-of-loop measurement of the SSB phase noise PSD of the $f_{\rm rep}$-stabilized comb 1, the SSB phase noise PSD of the THz wave ($L_{\rm THz}(f)$) at the 10-kHz frequency offset could be -105 dBc/Hz (see Fig. 5(b) and the supplementary material). The obtained phase noise is limited by the in-loop noise (\textgreater \ 10 kHz frequency offset), fiber noise and OSNR of the comb modes (\textless \ 10 kHz frequency offset). In addition to the careful handling of a long fiber, the use of widely-separated comb modes with high OSNR could improve the phase noise, in which injection locking of two CW lasers to the comb modes would be effective. In regard to measurement, the noise floor is more than 10 dB better than the obtained phase noise for the frequency offset of less than 100 kHz. When a smaller noise floor is required, the down conversion to a microwave from two THz waves can be realized by using a fundamental mixer, which improves the conversion efficiency. Although more progress would be required, the demonstrated system would be more integrated using Si photonics technologies. In particular, the rapid progress of the development of low loss, chip-scale waveguides would remove the use of a 100 m fiber \cite{puckett2021422, jin2021hertz}. The demonstrated system for the generation and characterization of low noise THz waves advances THz-based technologies for radars, wireless communications, and analog-digital converters in the era of 5G and 6G.

\section*{Methods}
\subsection*{Kerr comb generation}
An external cavity diode laser is used as a pump CW laser. The pump CW laser passes a dual-parallel Mach-Zehnder modulator (DP-MZM) for the fast scanning of the frequency of the pump CW laser, which is used to overcome the fast thermal dynamics of the microresonator induced when a chaotic comb transitions to a Kerr comb \cite{kuse2019control}. The DP-MZM is operated in a carrier-suppressed single sideband mode by applying an RF with a 90-degree phase difference to the two nested MZM and appropriately adjusting the phase bias. The frequency of a voltage-controlled oscillator (VCO, $f_{\rm vco}$) is controlled by a step function from an arbitrary waveform generator (AWG) to change the $f_{\rm vco}$. After generating the Kerr comb, the AWG is swapped to a DC power supply to reduce the phase noise of the VCO. Otherwise, an excess phase noise onto the pump CW laser is observed. The output from the DP-MZM is amplified by an EDFA, followed by an OBPF to reject the amplified spontaneous emission (ASE) from the EDFA. The pump CW laser with the optical power of 200 mW is coupled into a bus waveguide through a lensed fiber with an insertion loss of below 3 dB. The microresonator is made of Si$_3$N$_4$ and fabricated by the damascene process with the quality factor of more than \textgreater \ $1 \times 10^6$. After generating the Kerr comb, the residual pump CW laser is rejected by an optical bandstop filter, followed by an EDFA to amplify the optical power up to 160 mW. The amplified Kerr comb is split into two, which are directed to the setup for the TWDI and THz generation/characterization, respectively. The experimental setup is shown in the supplementary material. 
\subsection*{TWDI}
The Kerr comb is input into an i-MZM. Two AOMs (+80-MHz and -80-MHz shift, respectively) are used in both arms of the i-MZM to avoid the interference between the carrier from one arm and the residual carrier from the other arm. A long fiber (20, 80, 100, 110 m in the experiments) is installed. A PZT is attached onto a fiber to control the delay between the two arms, which is used to control the relative phase between two inputs of the mixer. The two outputs from the i-MZI pass through an OBPF to extract a pair of comb modes. In the experiments, the + 1st and -4th comb modes of comb 1 and +2nd and -3rd comb modes of comb 2 are used for the TWDI. The extracted comb modes are photodetected, generating RF signals with the carrier frequency of 160 MHz. After the RF signals are filtered and amplified, the RF signals are mixed, followed by a low pass filter to extract a DC signal, which shows the phase noise of the $f_{\rm rep}$. When the $f_{\rm rep}$ is stabilized, the frequency of the comb is modulated by controlling the pump current and PZT of the pump CW laser according to the error signal. The pump current is used for the fast, but narrow-range control, while the PZT is used for the slow, but wide-range control.
The transfer function of the iMZI used in the TWDI can be expressed as \cite{rubiola2005photonic}
\begin{equation}
    |H(if)|^2 = 4{\rm sin}^2(\pi \tau f)
\end{equation}
Here, $\tau$ is the time delay between the two arms of the i-MZI. The transfer function shows the scaling of $f^2$. 

\subsection*{THz generation and characterization}
Combs 1 and 2 are combined by an optical coupler. The four comb modes (two modes from each comb) are extracted by a waveshaper (4000s from Finisar), which are amplified by an EDFA to 32 mW. The power of the four comb modes before the waveshaper is carefully adjusted such that the power of the four comb modes after the EDFA become equal. The amplified comb modes are directed through a UTC-PD, generating two THz waves. Although the power of the two THz waves is not measured, power of each THz wave is likely to be around -26 dBm according to the datasheet. The two THz waves are collimated and focused onto a SBD, where the two THz waves are down-converted into a microwave through the square-law detection. The microwave is filtered by an RF low-pass filter and amplified by RF amplifiers. The phase noise of the microwave is measured by an electric spectrum analyzer. To measure the frequency of the microwave, the microwave is further down-converted by mixing with a local oscillator, followed by a low pass filter. The experimental setup is shown in the supplementary material.
\section*{Supplementary material}
See Supplement 1 for supporting content.


\bibliography{scibib}

\bibliographystyle{Science}

\section*{Acknowledgements}
G. N. acknowledges funding from the European Union’s Horizon 2020 research and innovation programme under the Marie Sklodowska-Curie grant agreement No 898074.

\subsection*{Data availability} The data that support the findings of this study are available from the corresponding author upon reasonable request.

\subsection*{Author contributions}
N. K. conceived the idea and performed the experiments with the assistance of K. N., Y. T., and S. O. G. N. and M. G. fabricated the Si$_3$N$_4$ microresonators. N. K. wrote the manuscript with contributions from all the authors. 
\subsection*{Competing interests}
The authors declare no competing interests.

\begin{figure}[h]
\centering
\fbox{\includegraphics[width=\linewidth]{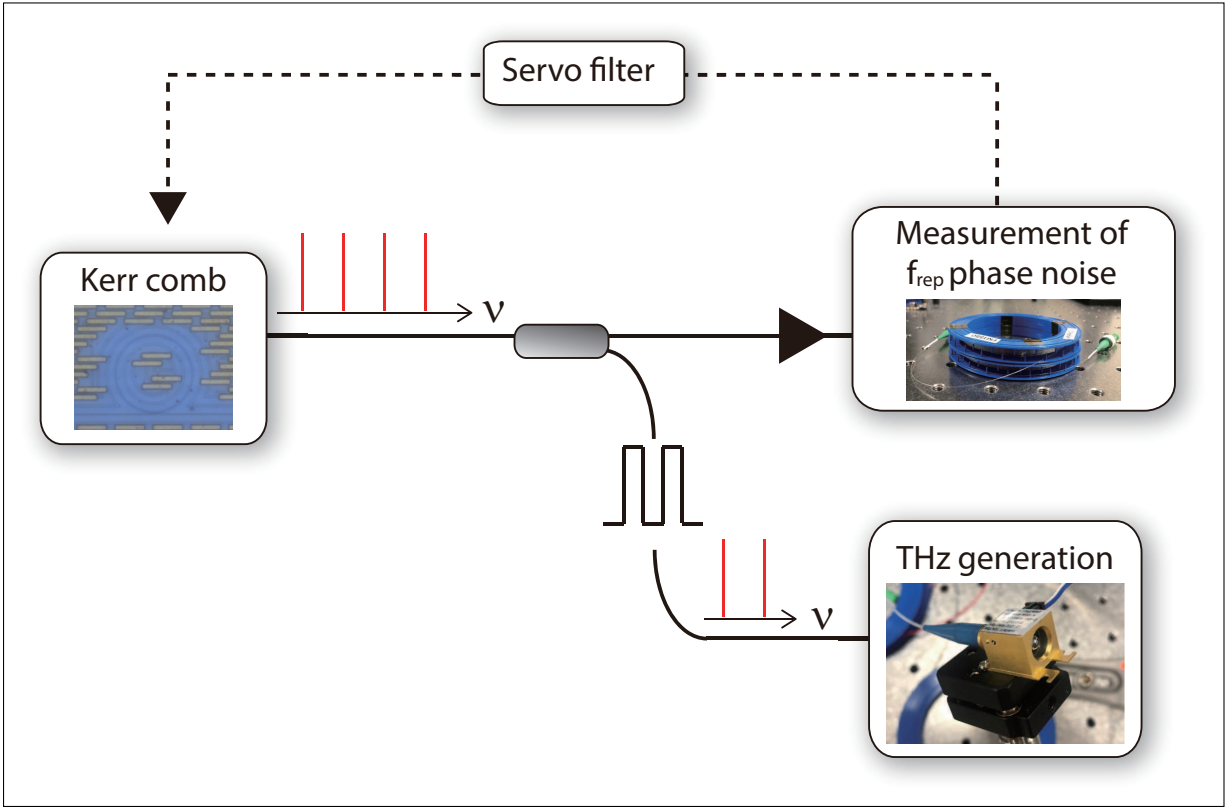}}
\caption{\textbf{Conceptual schematic of the generation of a low phase noise THz wave.} A Kerr comb is generated from a microresonator (see the photo). The phase noise of the $f_{\rm rep}$ of the Kerr comb is stabilized to a long fiber (see the photo) in a two-wavelength delayed self-heterodyne interferometer (TWDI) through a feedback loop. The two comb modes of the $f_{\rm rep}$-stabilized Kerr comb are extracted by an optical filter, followed by an uni-travelling-carrier photodiode (UTC-PD) (see the photo) to generate a THz wave. }
\end{figure}

\begin{figure}[h]
\centering
\fbox{\includegraphics[width=0.9\linewidth]{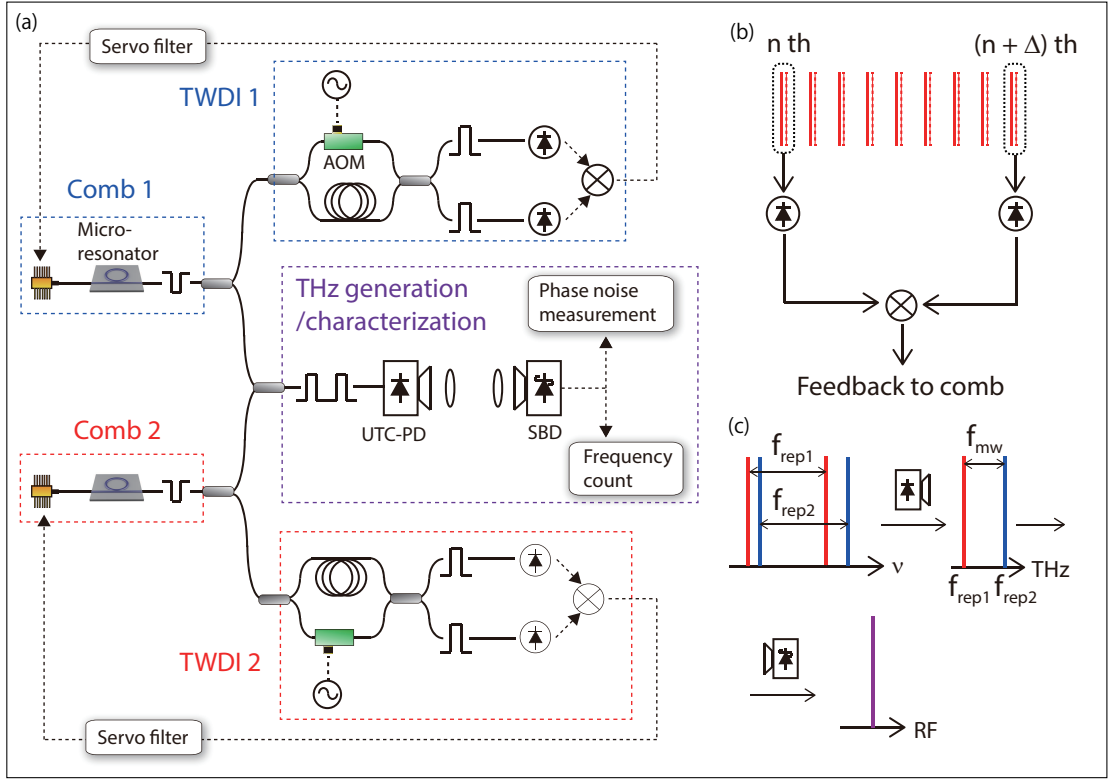}}
\caption{\textbf{Low phase noise THz genearation and characterization.} (a) Schematic of the experimental setup. The dashed blue (red) squares show the setups for comb 1 (comb 2) and TWDI 1 (TWDI 2). The dashed purple squares show the setup for the generation and characterization of a THz wave. AOM: acousto-optic modulator, SBD: Schottky barrier diode. (b) Illustration of the working principle of the TWDI. Pairs of single comb modes with (solid line) and without (dotted line) delay are photodetected, generating signals with the phase noise PSDs of $|H(if)|^2\{L_{\rm ceo}(f) + n^2L_{\rm rep}(f)\}$ and $|H(if)|^2\{L_{\rm ceo}(f) + (n+\Delta)^2L_{\rm rep}(f)\}$. By mixing the two signals, a DC signal with the phase noise PSD of $|H(if)|^2\cdot \Delta^2 L_{\rm rep}(f)$ is generated, which is used as an error signal. (c) Illustration of the working principle of the characterization of the phase noise of the THz. Four comb modes from combs 1 and 2 are directed into an UTC-PD, generating two THz waves with the carrier frequencies of $f_{\rm rep1}$ and $f_{\rm rep2}$. The two THz waves are squared-law-detected at a SBD, generating a microwave with the phase noise of the sum of the two THz waves.}
\end{figure}

\begin{figure}[h]
\centering
\fbox{\includegraphics[width=\linewidth]{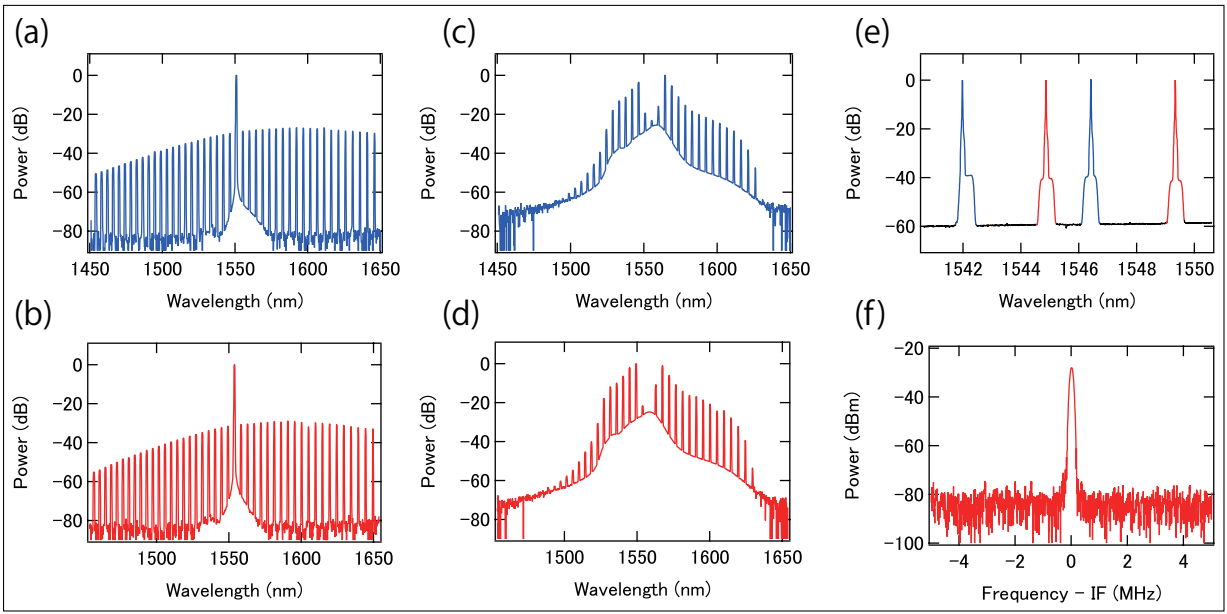}}
\caption{\textbf{Optical spectra and RF spectrum.} Optical spectra of combs 1 (a) and 2 (b) before the bandstop filter and amplified optical spectra of combs 1 (c) and 2 (d) after the bandstop filter. (e) Optical spectrum of four comb modes (blue curve from comb 1 and red curve from comb 2) used for THz waves generation. (f) RF spectrum of the microwave generated from the SBD. Resolution bandwidth (RBW) for (a) - (d) is 1 nm. RBW for (e) is 0.02 nm. RBW for (f) is 100 kHz.}
\end{figure}

\begin{figure}[h]
\centering
\fbox{\includegraphics[width=\linewidth]{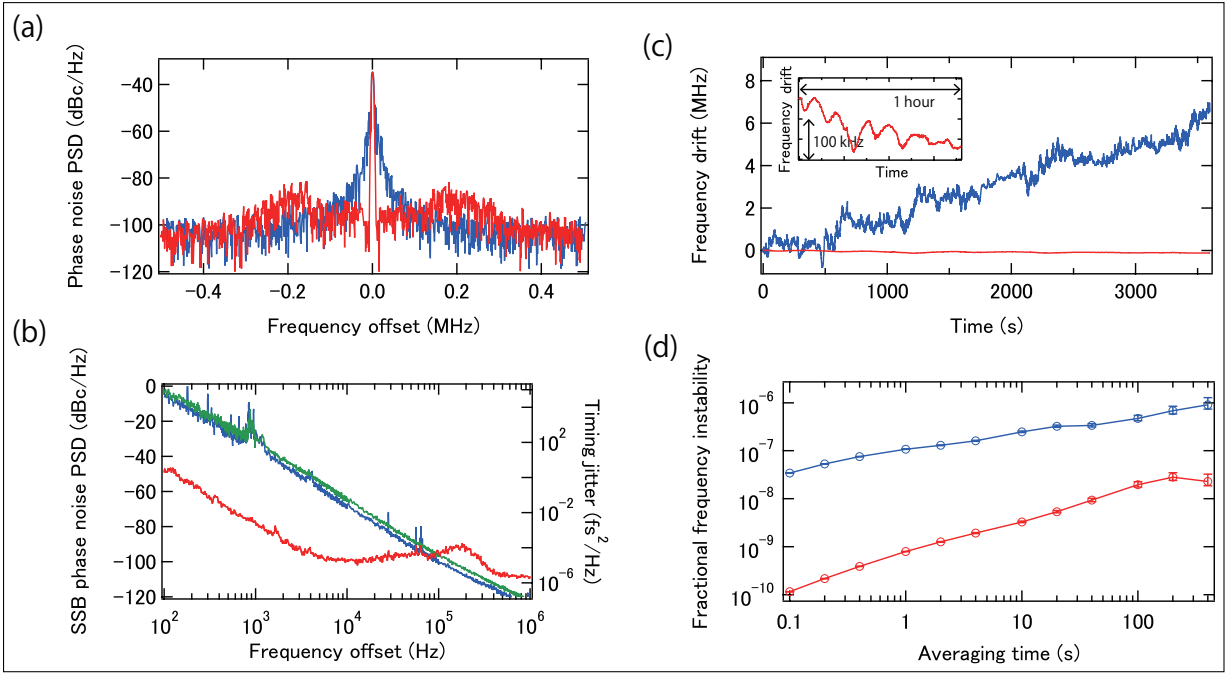}}
\caption{\textbf{Phase noise spectra and frequency instability of the THz wave.} (a) RF spectra of the down-converted microwave with (red curve) and without (blue curve) locking, measured with the RBW of 3 kHz. (b) Single-sideband (SSB) phase noise power spectral density (PSD) and timing jitter of the down-converted microwave (red curve) and that of the repetition frequency of free-running combs 1 (blue curve) and 2 (green curve) measured by the TWDI. (c) Frequency drift of the down-converted microwave with (red curve) and without (blue curve) locking. The inset shows the magnified frequency drift of the down-converted microwave with locking. (d) Frequency stability of the down-converted microwave calibrated at 560 GHz with (red line and circle) and without (blue line and circle) locking.}
\end{figure}

\begin{figure}[h]
\centering
\fbox{\includegraphics[width=0.9\linewidth]{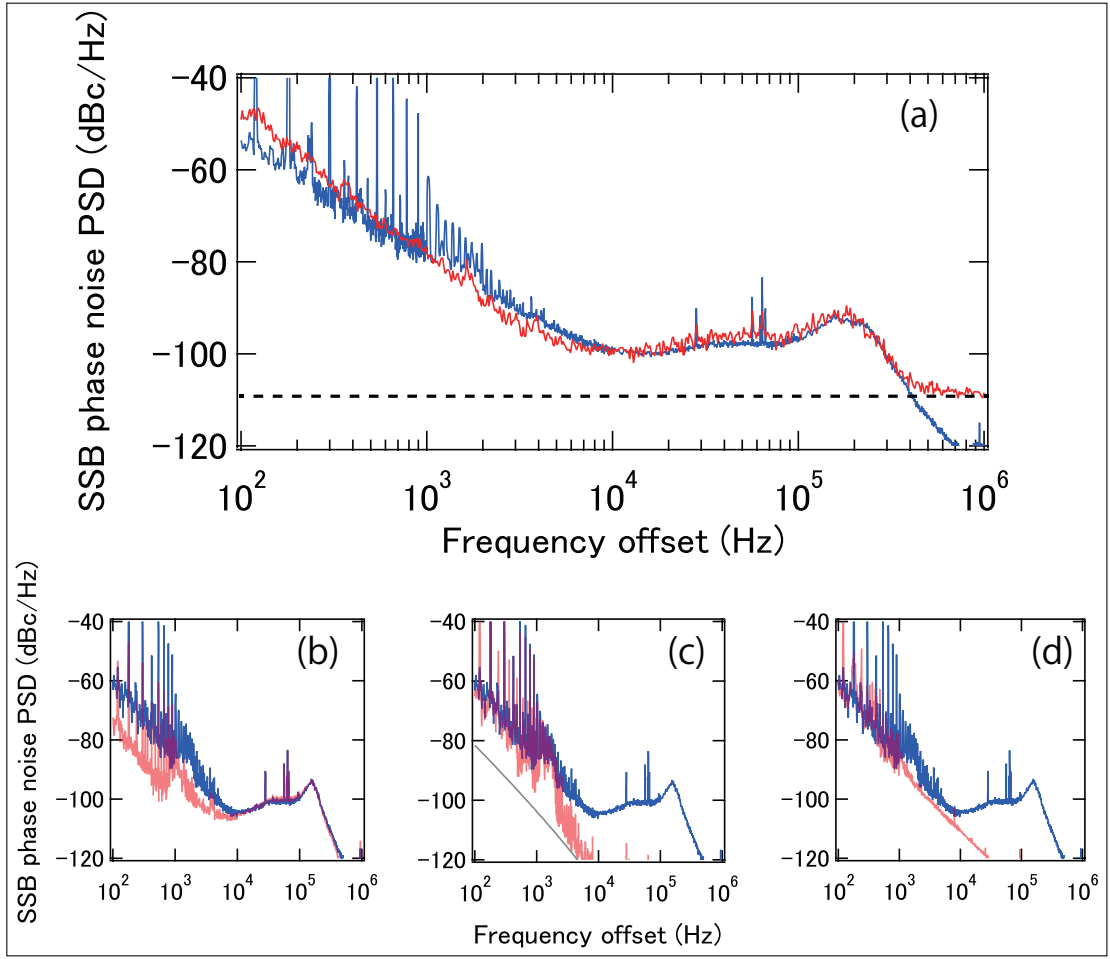}}
\caption{\textbf{Phase noise limit.} (a) SSB phase noise PSDs of the microwave down-converted from the two THz waves (red curve) and the sum of the out-of-loop phase noise PSDs of $f_{\rm rep1}$ and $f_{\rm rep2}$ (blue curve). The dashed line shows the measurement noise floor for the microwave limited by the generated THz power. (b) In-loop phase noise PSD of the stabilized $f_{\rm rep}$ of comb 1 (red curve). (c) Fiber noise limit (red curve) and theoretical thermal noise of the fiber \cite{wanser1992fundamental, duan2012general,dong2016observation} (gray line). (d) Optical signal-to-noise ratio (OSNR) limit of the TWDI (red curve). The blue curves in (b), (c), and (d) are the out-of-loop phase noise PSD of the stabilized $f_{\rm rep}$ of comb 1.} 
\end{figure}

\begin{figure}[h]
\centering
\fbox{\includegraphics[width=0.6\linewidth]{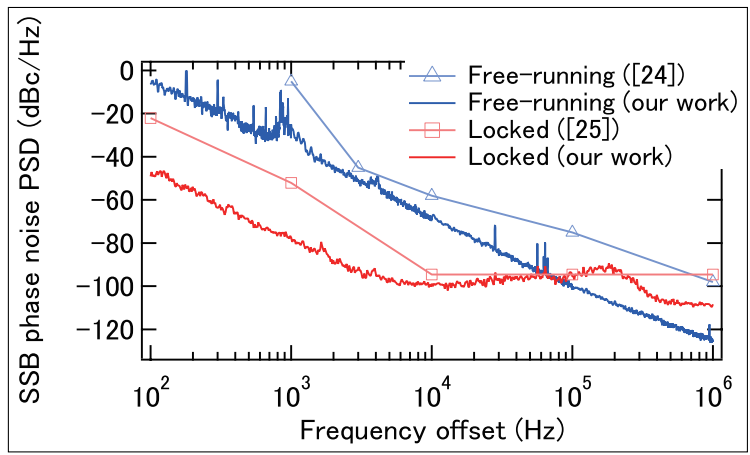}}
\caption{\textbf{Comparison of SSB phase noise PSDs scaled to 560 GHz.} SSB phase noise PSDs of 560-GHz carriers generated from free-running Kerr combs reported in the reference \cite{wang2021towards} (light blue triangle) and this work (red square). SSB phase noise PSDs of 560-GHz carriers generated from stabilized Kerr combs reported in the reference \cite{tetsumoto2021optically} (light red curve) and this work (red curve). The phase noise shown by the blue curve is estimated from the measurement by a TWDI, instead of actually generating a 560-GHz carrier. When a 560-GHz carrier is generated, phase noise will see the white noise floor similar to the red curve. The red curve is sum of the two comb ($L_{\rm rep1}(f)$ and $L_{\rm rep2}(f)$). Therefore, the phase noise of a single 560-GHz carrier would be a few dB better.}
\end{figure}



\clearpage

\end{document}